\documentclass[10 pt,final,doublecolumn]{Latex_styles/IEEEtran}

\usepackage{epsfig}
\usepackage{subfigure}
\usepackage{amsopn,amsmath,amssymb,amsfonts}
\usepackage{cite}
\usepackage{Latex_styles/citesort}
\usepackage{balance}
\usepackage{float}
\usepackage{graphics}
\usepackage{array}
\usepackage{multirow}
\usepackage{algorithm,algorithmic}
\usepackage{xcolor}
\usepackage{breqn}

\begin{document}

\title{System Modelling and Design Aspects of Next Generation High Throughput Satellites}

\author{
Shree Krishna Sharma, Jorge Querol, Nicola Maturo, Symeon Chatzinotas, and Bj$\ddot{\mathrm{o}}$rn Ottersten 
\thanks{\footnotesize{The authors are with the SnT, University of Luxembourg, Luxembourg 1855, Luxembourg. Email: \{shree.sharma, jorge.querol, nicola.maturo, symeon.chatzinotas, bjorn.ottersten\}@uni.lu}.}
\thanks{\footnotesize{This work was supported in part by FNR, Luxembourg under the CORE projects 5G-Sky, ASWELL, FLEXSAT, DISBUS and PROCAST.}}}

\maketitle

{}

\begin{abstract}
Future generation wireless networks are targeting the convergence of fixed, mobile and broadcasting systems with the integration of satellite and terrestrial systems towards utilizing their mutual benefits. Satellite Communications (SatCom) is envisioned to play a vital role to provide integrated services seamlessly over heterogeneous networks. As compared to terrestrial systems, the design of SatCom systems require a different approach due to differences in terms of wave propagation, operating frequency, antenna structures, interfering sources, limitations of onboard processing, power limitations and transceiver impairments. In this regard, this letter aims to identify and discuss important modeling and design aspects of the next generation High Throughput Satellite (HTS) systems. First, communication models of HTSs including the ones for multibeam and multicarrier satellites, multiple antenna techniques, and for SatCom payloads and antennas are highlighted and discussed. Subsequently, various design aspects of SatCom transceivers including impairments related to the transceiver, payload and channel, and traffic-based coverage adaptation are presented. Finally, some open topics for the design of next generation HTSs are identified and discussed.
\end{abstract}

\IEEEpeerreviewmaketitle
\vspace{-10 pt}
\section{Introduction}
\label{sec: sec1}
Satellite Communication (SatCom) is considered as an important segment of future 5G and beyond wireless networks, as it can provide several benefits including broadcasting capability, ubiquitous coverage and broadband connections to inaccessible/remote areas. In addition to complementing terrestrial wireless connectivity in several ways, SatCom is better suited for novel 5G and beyond applications such as content delivery networks and distributed Internet of Things (IoT) networks, and is a viable solution to provide telecommunication services to a wide range of areas including communications-on-the-move and high-speed platforms (i.e., airplanes, unmanned aerial vehicles), as well as emergency rescue and disaster relief scenarios \cite{5GbookIET}.

Recent advances in high frequency (Ku-band, Ka-band, EHF-band, optical) technologies, digital payload and signal processing have led to the emergence of High Throughput Satellite (HTS) systems \cite{Perez2019}. Furthermore, several promising techniques and paradigms including digital twin, reconfigurable onboard processors,  Software-Defined Networking (SDN),  Network Function Virtualization (NFV) and network slicing aim to further make future HTSs more flexible and dynamic towards supporting  non-uniform and rapidly time varying traffic demands across multiple beams and diverse Quality of Service (QoS) requirements of emerging services \cite{5GbookIET}. Although satellite systems have moved from the conventional monobeam scenario to the multibeam platform, cochannel interference issues caused by full-frequency reuse needs to be addressed by applying advanced precoding and multiuser detection (MUD) schemes. Besides, as the number of Geostationary (GSO) and Non-GSO (NGSO) satellites is increasing over the recent years,  the need of coexistence of satellite systems with the terrestrial wireless systems has become a necessity. Moreover, several challenges in terms of enhancing system capacity, spectral efficiency and coverage,  meeting latency and reliability requirements, and mitigating transceiver impairments need to be addressed in order to effectively integrate SatCom with 5G and beyond wireless networks.

In the above context, this letter aims to provide system  modelling and design guidelines of the next generation HTS systems. First, various communication models of HTSs including the ones for multibeam satellites, multicarrier SatCom, multi-antenna techniques and payload models will be discussed (Sec. \ref{sec: sec2}). Secondly, several design aspects of SatCom transceivers including transceiver and payload impairments, Channel State Information (CSI) related impairments and traffic-aware coverage adaptation will be described (Sec. \ref{sec: sec3}). Finally, open research topics for the next generation HTSs will be briefly discussed (Sec. \ref{sec: sec4}).
\vspace{-10 pt}
\section{Communication models for HTSs}
\label{sec: sec2}
\subsection{Models for Multibeam satellites}
\subsubsection{Frequency Reuse in Multibeam Satellites}
Like in cellular networks, the term ``frequency reuse'' in multibeam satellites refers to the reuse of user link bandwidth across multiple beams of a satellite. As compared to the widely-used  four color reuse method in the conventional multibeam satellites, the trend is moving towards full frequency reuse, however, this results in a high level of cochannel interference, leading to the need of advanced precoding and MUD schemes.

Let $K$ denote the frequency reuse factor and total available bandwidth in the forward link is $B$, then the $i$-th user beam ($B_i$) can be written as: $B_i=B/K=N_iB_c/K$, with $B_c$ being the carrier bandwidth, and $N_i$ the number of carriers in the $i$-th beam. Then, the system throughput of a multibeam system is given by \cite{Rgm:12}: 
$C=B/K \sum_{i=1}^{N_b} \mathrm{log}_2(1+\gamma_i)$,  where $N_b$ is the number of beams, and $\gamma_i$ denotes the Signal to Interference plus Noise Ratio (SINR). The lower value of $K$ results in higher available bandwidth per beam but also increases the co-channel interference. As compared to the regular frequency reuse pattern and uniform carrier/power allocation, future flexible multibeam satellites are expected to support non-regular frequency reuse pattern and non-uniform power/carrier allocation. 
\subsubsection{Beamhopping Multibeam System} In a beamhopping system, the available bandwidth is reused  within a cluster in the time domain instead of only frequency reuse in the conventional multibeam systems \cite{Sharmacogbeamhop}. The employed beamhopping technique can utilize either full frequency or partial frequency reuse depending on whether all the available bandwidth or its segment is allocated to each illuminated beam. Let $N_t$ denote the number of time slots in each time window, then the beamhopping pattern can be characterized with an illumination matrix $\mathbf{T}$ of the size $N_{t} \times N_{b}$, with its element $T_{i,j} \in \{0,1\}$ indicating whether the $j$-th time slot is allocated to $i$-th beam or not,  and the total number of time slots assigned to the $i$-th beam can be written as: $N_{i,t}=\sum_{j=1}^{N_t} T_{i,j}$.

\subsubsection{Multiple Smart Gateways (GWs)} 
Towards addressing the limitations of feeder link bandwidth in multibeam systems, multiple GWs approach seems promising as it can enable the operation of feeder links in Q/V bands by ensuring a high-level of site diversity \cite{Kyrgiazos2014gateway}. However, there arise the issues of intrasystem interference, and complexity in employing advanced transmission techniques (i.e., precoding, beamhopping). To address these, the concept of smart GW diversity is emerging, in which each user beam is served by a number of GWs deployed in different geographical locations instead of a single GW, and they are interconnected via high-speed terrestrial links. With this solution, link availability is significantly enhanced as the traffic from the GW experiencing deep fades can be routed to the un-impacted GWs located in different locations.   
Instead of making all GWs active, $P$ redundant per $N$ active GWs can be utilized to minimize the number of GWs from $2N$ to $P+N$. 
With a user terminal being served by $N$ active GWs and $M$ number of users, the offered capacity to beam $j$ in the cases of frequency multiplexing and time multiplexing are respectively given by \cite{Kyrgiazos2014gateway}; $B_j=\sum_{i=1}^N C_{i,j}^F$ and $B_j=\sum_{i=1}^N C_{i,j}^T X_{i,j} T_s$, with  $i=1,2,....N$, $j=1,2....M$,  where $C_{i,j}^F$ and $C_{i,j}^T$ denote the instantaneous average offered capacity from the $i$-th GW to the $j$-th user beam, respectively, and $X_{i,j} $ denotes the number of time slots that the $i$th feeder link is connected with the $j$th user link and $T_s$ denotes the slot duration. 

\subsubsection{Multibeam Joint Processing (MJP)} Like multicell joint processing used in terrestrial cellular systems to mitigate interferences and enhance the system capacity, MJP can be employed in multibeam satellites by jointly processing multiple users with the help of multiuser precoding and joint decoding at the forward and return links, respectively \cite{SharmaIA2013,Christopoulos2012}. With this MJP approach, the interference channels of forward and return links can be realized with Multiple Input Multiple Output (MIMO) broadcast  and Multiple Access Channel (MAC), respectively \cite{Christopoulos2012}. 

Considering a cluster of $K$ beams supporting $K$ user terminals equipped with a single antenna,  
the input-output equation for the k-th beam can be written as: $y_k=\sum_{i=1}^Kh_{k,i}x_i+z_k$, where $h_{k,i}$ being the complex channel coefficient between the $k$-th beam and the $i$-th user, and $z_k$ is the Additive White Gaussian Noise (AWGN) at the receive antenna. In general, the baseband model for all the beams can be written as: $\mathbf{y}=\mathbf{H}\mathbf{x}+\mathbf{z}$, where $\mathbf{y}$, $\mathbf{x}$ and $\mathbf{z}$ are $K \times 1$ vectors of the received signal, transmit signal and AWGN, respectively, and $\mathbf{H}$ denote the $K \times K$ channel matrix, whose modeling should consider various aspects of a satellite channel including beam gain, Rician fading, lognormal shadowing and antenna correlation.
\vspace{-10 pt}
\subsection{Multicarrier Satellite Systems}	
\vspace{-5 pt}
A multicarrier satellite system includes $M$ number of independent carriers, with each carrier employing forward error correction coding, and then followed by interleaving and Gray mapping onto a higher-order modulation constellation with the alphabet size of $M$. Thus generated composite signal in the complex-valued form is given by \cite{5GbookIET};  $S(t)=\sum_{m=1}^M \frac{1}{\sqrt{M}}. s_m(t). e^{j(2\pi f_mt+\theta_m)}$, where $f_m$ is the m-th carrier frequency, $s_m(t)$ is the transmitted signal at the $m$th carrier at the $t$th time instant and $\theta_m$ denotes the normalized difference in the carrier phase. Subsequently, on the transponder, the signal goes through an IMUX (input multiplexer) filter, a non-linear High Power Amplifier (HPA) and an OMUX (output multiplexer) filter, and thus generated non-linear channel with memory can be modeled utilizing the Volterra series \cite{Beidas2011}.

In multicarrier SatCom systems, interference may occur between adjacent carriers, and can be modeled as a function of the number of subcarriers and bandwidth compression factor. Assuming that each carrier goes through independent flat-fading channels,  the multi-carrier channel matrix $\mathbf{H}$ for $M$ number of carriers can be written as \cite{Freqpacking2013VTC}:
$\mathbf{H}=\left[
            \begin{array}{cccc}
              h_1 & \mu h_2 & \ldots & 0 \\
              \mu h_1 & h_2 & \ldots & 0 \\
              \vdots & \vdots  & \vdots & \vdots \\
              0 & 0  & h_{M-1} & \mu h_M \\
               0 & 0 & \mu h_{M-1}  & h_M \\
            \end{array}
               \right]$,
where the correlation amplitude $\mu$ characterizes the effect of intercarrier interference, and the parameter $h_i$ denotes the Rician fading channel coefficient, given by; $h= \left( \sqrt{\frac{K_r}{K_r+1}} l+\sqrt{\frac{1}{K_r+1}} g \right)$, where \(K_r\) is the Rician factor, $l$ represents the deterministic LoS component and $g$ denotes the Rayleigh fading coefficient.
\vspace{-10 pt}
\subsection{Multi-antenna Techniques for SatCom}
Multi-antenna beamforming (BF)/precoding in SatCom systems can provide the benefits of enhanced system capacity and the mitigation of inter-beam interference towards realizing multibeam and multi-spot transmissions.  
\subsubsection{Beamforming for SatCom} As compared to the terrestrial MIMO scenarios, digital BF using an array of multi-beam antennas differs mainly in terms of multi-user diversity being difficult due to the involved line of sight component and limited channel dynamics, and the antenna design as it impacts the geometric coverage of SatCom systems \cite{Zhang2019multiple}. The performance of a digital beamformer is mainly characterized with the (i) antenna structure, and (ii) design of BF weights. As compared to the widely-used Uniform Linear Array (ULA) structure in terrestrial BF design, antenna structure at the Satellite terminals is mainly an Array Fed Reflector (AFR). The response vector $\mathbf{a} (\phi,\theta)$ of an AFR is given by \cite{Sharma3D2015}; $\mathbf{a} (\phi,\theta)=[g_{1} e^{j \Psi_{1}},g_{2} e^{j \Psi_{2}},\ldots,g_{M} e^{j \Psi_{M}}]^T$, where $g_{i}$ and $\Psi_{i}$ denote the amplitude gain and the phase of the $i$th feed ($i=1,\ldots,M$) to a unit amplitude plane wave coming from the direction ($\phi,\theta$), respectively. The  $M \times 1$ received signal vector at the satellite terminal equipped with $M$ number of multiple Low-Noise Block downconverters (LNBs) while considering the desired signal $s_0$ from the desired FSS satellite located at the direction of  $\phi_0,\theta_0)$,  $K$ number of interfering co-channel terrestrial Fixed Service (FS) stations can be written as:
  $\mathbf y = h_0 \mathbf a(\phi_0,\theta_0)  s_0+\sum_{k=1}^K h_k \mathbf{a} (\phi_k,\theta_k) s_k+\mathbf{z}$,
where  $a(\phi,\theta)$ denotes the array response vector in the direction of $(\phi, \theta)$, $s_k$ is the transmitted signal from the $k$th interfering FS station, $h_k$ represents the channel gain from the $k$-th station to the satellite terminal, and $\mathbf{z}$ denotes AWGN vector. Then, the output of the beamformer is obtained by linearly combining the received signal vector $\mathbf{y}$ with an $M \times 1$ complex weight vector $\mathbf{w}$ as: $y_1=\mathbf{w}^{\dagger} \mathbf y$, where $(\cdot)^{\dagger}$ denotes the Hermitian transpose. The BF weights $\mathbf{w}$ can be designed using suitable BF techniques such as Capon, Linearly Constrained Minimum Variance or optimization-based techniques to maximize SINR or minimize total power, depending on the desired performance objective.

\subsubsection{Precoding for SatCom} The main objectives of precoding are to enable multi-stream communications towards maximizing the link throughout at the receiver, and  to counteract various types of impairments including different interferences (multiuser, adjacent channel, inter-symbol) and non-linear effects caused due to hardware imperfections, assuming they are known or can be modeled at the Tx \cite{Perez2019}.
Considering a broadband multibeam satellite serving $K$ numbers of users and having $M$ number of feeds,  and single user per beam scheduled in each time slot, the $K \times 1$ signal vector received by $K$ users can be written as: $\mathbf{y}=\mathbf{H}\mathbf{W}\mathbf{x}+\mathbf{z}$, where $\mathbf{x}$ is the $K \times 1$ signal vector to be transmitted to $K$ users, $\mathbf{W}$ is $M \times K$ is the precoding matrix, $\mathbf{H}$ is $K \times M$  channel matrix of the considered multibeam satellite channel, and $\mathbf{z}$ denotes the $K \times 1$ AWGN vector. The multibeam channel matrix $\mathbf{H}$ can be written as \cite{Qiprecoding2018}: $\mathbf{H}=\mathbf{\Phi} \mathbf{B}$, with the $K \times K$ matrix $\mathbf{\Phi}$ denoting the phase variations caused due to propagation effects,  and $K \times M$ matrix $\mathbf{B}$ being the multibeam antenna pattern, which depends on various parameters including the gain between a particular feed and the user of interest, receive antenna gain, distance between satellite and the user, frequency of operation and bandwidth. The precoding matrix $\mathbf{W}$ should be designed to meet the desired performance objectives such as energy efficiency, sum-rate maximization and fairness. 
\vspace{-10 pt}
\subsection{Models for SatCom Payloads and Antennas}
\vspace{-5 pt}
\subsubsection{Payload types}
Satellite payloads can be broadly classified into two groups, namely, regenerative and repeaters, and repeaters can be further subdivided based on whether signal can be processed in the digital domain or not \cite{SpaceAntennaHandbook2012}. Based on this, three most commonly used payload types are the following: i. \textit{Regenerative Transponder}: This payload receives, demodulates, processes, re-modulates, and re-transmits the signal. This is the most complex type of payload as it requires a full Tx and Rx chain for each transponder. 
ii. \textit{Digital Transponder}: In this type, the received signal is digitally processed at some blocks of the path chain including  channelization, signal routing, digital filtering, or the programmable gain amplifier. This payload offers better power efficiency and flexibility than its analog counterpart. 
iii. \textit{Bent-Pipe Transponder}: In this payload, uplink signals are just amplified, filtered, frequency translated and routing via a switching matrix, entirely with analog components. This is a widely used configuration in the current in-orbit satellites due to its simplicity and reliability, however, the recent trend is to migrate to digital transponders with mixed analog and digital components. 
\subsubsection{Antenna models}
The state-of-the-art of satellite antennas is clearly dominated by the passive reflector antennas, with very efficient and optimized designs developed during the years \cite{SpaceAntennaHandbook2012}. For multi-beam pattern generation, two types of reflector based configurations, namely, Single Feed Per Beam (SFPB) and Multiple Feed Per Beam (MFPB) configurations are utilized. Despite the widespread use of passive reflectors, the emerging trend is to incorporate active antenna arrays in future comm. satellites to address the demanding requirements in terms of pattern flexibility, power and frequency reconfigurations, electronic beam steering and meeting non-uniform traffic demands.

\vspace{-10 pt}
\section{Design aspects of SatCom Transceivers}
\label{sec: sec3}
\subsection{Transceiver and Payload Impairments}
In any communications system, there exist several transceiver and channel induced impairments that degrade the signal and, thus, the overall performance of the system. Besides, there are some impairments that are more specific to SatComs and these are the ones treated henceforth.

\subsubsection{Non-linearity Effects} Satellite payload impairments are, obviously, the most representative ones of SatCom. Most satellite HPAs are based on Traveling-Wave Tube Amplifier (TWTA) technology. HPAs are driven close to their saturation points for power efficiency, thus, TWTA operation introduces some non-linearity effects related to the Amplitude Modulation (AM) and the Phase Modulation (PM) responses. Thus, the non-linearity effects are represented by AM/AM and AM/PM responses. The relation between the baseband equivalent input and output of the HPA can be expressed using a polynomial model of order $2J+1$ as \cite{Delamotte2016}: $y(t)=\left[\sum_{j=0}^{J}\gamma_{2j+1}\frac{1}{2^{2j}}\binom{2j+1}{j}\vert x(t)\vert^{2j}\right]x(t)$, where $\gamma_{2j+1}$ are the coefficients representing AM/AM and AM/PM responses.

\subsubsection{Multicarrier distortion} The joint amplification of multiple signal carriers in the same HPA is a cost-effective solution. In current and future designs, the use of a single HPA per carrier is not feasible due to the unreasonable increase of weight and size \cite{Delamotte2016}. However, multicarrier operation generates strong non-linear intermodulation distortions of the amplified signals. As highlighted earlier, Volterra series is a well-known mathematical tool for the modeling of non-linear systems \cite{Beidas2011}. 

\subsubsection{Frequency offset and Phase Noise (PN)}
Frequency offset arises due to the fact that the Tx and the Rx oscillators are different and placed in physically separated locations and, therefore, their fundamental oscillation frequencies become different. This frequency offset is usually modeled as the difference between the Tx and Rx oscillators, $\phi_t = 2\pi f_t t + \varphi_t$ and $\phi_r = 2\pi f_r t + \varphi_r$. Hence, the received down-converted signal $r(t)$ can be expressed as: $r(t) = s(t)\,e^{-j2\pi f_n t + \varphi_n}$, where $f_n$ is the frequency offset and $\varphi_n$ is the phase offset between the oscillators.

The PN is a generalization of the frequency offset concept, which takes into account the variations of the Tx oscillator phase $\phi_t$ and the Rx oscillator phase $\phi_r$. PN emulation can be carried out based on the superposition of multiple characteristics of the power spectral density $S_\phi(f) = \sum_{\alpha = 0}^4 h_\alpha / f^\alpha$ \cite{PhaseMartinez}, where the terms are related to random walk FM, flicker FM, white FM, flicker and white PNs respectively.

\subsubsection{Doppler effect}
Doppler effect appears when the relative velocity vector $\overrightarrow{v_d}$ between the Tx and Rx is different than zero. In SatComs, LEO satellites can address the problem of GEO large delays in the communication link to a certain extent, however, they suffer from an increased Doppler shift $f_d$. The relationship between the Doppler frequency shift and the relative velocity is given by; $f_d = f_0\,\overrightarrow{v_d}/c$, where $f_0$ is the carrier frequency, $c$ the speed of light and $v_d = d(P_s - P_t)/dt$ is obtained in spherical coordinates with $P_s$ and $P_t$ the position of satellite and Earth transceiver, respectively. A closed-form solution of $f_d$ with respect to the time $t$, when the satellite is at the maximum elevation angle can be expressed as \cite{DopperKhodeli}:
$f_{d}(t) = -\displaystyle \frac {f_0}{c}\cdot \frac {w_{s}r_{e}r_s\sin(w_{s}t)\,\eta(\theta _{\max })}{\sqrt {r_{e}^{2}+r_s^{2}-2\,r_{e}r_s\cos(w_{s}t)\,\eta(\theta _{\max })}}$,
where
$\eta(\theta_{max}) = \cos \left[\cos^{-1}\left({\frac{r_{e}}{r_s}} \cos(\theta_{max})\right)-\theta_{max}\right]$,
$f_0$ is the carrier frequency, $w_s$  is the angular velocity of the satellite in the Earth central inertial frame, $r_e$ is the radius of Earth, $r_s$ is the satellite's orbit radius, and $\theta_{\max}$ is the maximum elevation angle. 

In the frequency domain, the center frequency of the received signal $r(t)$ as \cite{DopplerZhao}: $r(t) = s(t)\,e^{-j2\pi\left(f_0 + f_d(t)\right)t + \theta_0} + n(t)$, where $s(t)$ is the base-band  signal from the Tx, $\theta_0$ is the phase offset of the up-converter oscillator, and $n(t)$ is AWGN. Whereas, in the time domain, Doppler effect produces compression or stretching of the signal, which can be modelled as a change in the sampling rate of the received discrete frequency-compensated signal as: $r[kT_s] = s[k(1 + f_d(t)/f_0)T_s]$, where $k$ is the sample index and $T_s$ is the symbol time.

\subsubsection{I/Q Imbalance}
The I/Q imbalance is an impairment present during the I/Q up- or down-conversion of the complex base-band signal \cite{IQImXu}. Considering the signal before the converter $x(t) = x_I(t) + jx_Q(t)$, the signal at its output can be expressed as $y(t) = (1 + \epsilon_A)\left[x_I(t) \cos(\epsilon_\phi/2) - x_Q(t) \sin(\epsilon_\phi/2)\right] + j(1 - \epsilon_A)\left[x_Q(t) \cos(\epsilon_\phi/2) - x_I(t)\sin(\epsilon_\phi/2)\right] = \eta\,x(t) + \eta'\,x^*(t)$, where $\epsilon_A$ and $\epsilon_\phi$ are the I/Q amplitude and phase imbalance, respectively. Note that if $\epsilon_A = 0$ and  $\epsilon_\phi = 0$, then $\eta = 1$ and $\eta' = 0$.  The principal effect of the I/Q imbalance can be observed as a signal image in the frequency domain since $Y(f) = \eta\,X(f) + \eta'\,X^*(-f)$, where $X(f)$ and $Y(f)$ are the Fourier transforms of $x(t)$ and $y(t)$, respectively.
\vspace{-10 pt}
\subsection{CSI acquisition and channel impairments}
The widely-used assumption of perfect CSI, either statistically or instantaneously, is rather impractical due to various inevitable channel impairments such as imperfect channel estimation, limited feedback, or latency-related errors. In the case of imperfect channel knowledge, the estimated channel matrix ${\hat{\bf H}}$, which along with the channel-error matrix ${\bf E}$ can be written as: ${\hat{\bf H}}={\bf H}+{\bf E}$, where the elements of the channel-error matrix ${\bf E}$ are independent and identically Gaussian-distributed with zero mean and variance $\sigma^2_E$ of real and imaginary parts. A more complete model is obtained if the CSI delay is considered.  Assuming the channel to be constant for one symbol time and a delay of $D$ symbol time, ${\hat{\bf H}}$ can modeled as \cite{BasharCSI}: $\hat{\bf H}[n]={\bf H}[n-D]+{\bf E}[n]$ where $\hat{\bf H}[n]$, ${\bf H}[n]$, ${\bf E}[n]$ denote the estimated channel, true channel and error matrices at the n-th time instance, respectively. The correlation coefficient $\rho$ is obtained based on the classical Clarke's isotropic scattering model given by \cite{OnggosanusiCSI} ; $\rho = J_0\left(2\pi f_d D T_s\right)$, $f_d$ denotes Doppler shift, $J_0(\cdot)$ is the zeroth-order Bessel function of the first kind, $T_s$ denotes one symbol time and $D$ is always considered to be a non-negative integer. The variance of the error vector is related to $\rho$ as: $\sigma^2_E = 1 - \rho^2$.
\vspace{-15 pt}
\subsection{Traffic-aware coverage adaptation}
Due to very high initial costs, traditional satellites are generally designed for a long span, but while this approach fits fine with broadcast services, in case of broadband transmissions, any fixed multibeam pattern and footprint design may not be well suited because of the high dynamicity of the broadband demand. 
However, most existing works focus on having beams of equal size and ensure global coverage but have not considered the user demand during the beam pattern design. Only recently new footprint design techniques have been started to be developed that take into account users' locations and traffic demand in order to optimize the capacity distribution across the beams. The basic idea of these approaches is to try to design the beam footprint taking into account the geographical distribution of the users in the coverage area, in order to guarantee more or less uniform distribution of the traffic demand across the beams. The process is summarized in the following steps: 1) Grouping of $N$ users into well defined clusters with equal traffic distribution, 2) Tessellation of the coverage area, 3) Assignment of a beam to each cluster, and 4) Derivation of the beam pattern and the satellite antenna gains in the different users' locations.

To accomplish the first step, different solutions are already available in literature \cite{XU2015}, such as k-means, k-medoids, Partitioning Around Medoids (PAM) and Clustering LARge Applications (CLARA). Step 2 is needed in order to guarantee the absence of areas with no coverage. This is particularly important when mobile users are considered in the pool (e.g., ships, airplanes). In Steps 3 and 4, the actual beam pattern is generated according to the clusters and the tessellation defined in Steps 1 and 2. Clearly, this process can be repeated each time that a significant shift in the traffic demand happens in order to update the beam pattern accordingly with the latest traffic demand requirements.
\vspace{-10 pt}
\section{Open topics}
\vspace{-5 pt}
\label{sec: sec4}
Herein, we briefly highlight some important open topics for future HTSs. For further topics, interested readers may refer to \cite{Shankar2020comm,Kisseleff2020}.  
\vspace{-5 pt}
\subsubsection{Onboard Processing (OBP)} 
The reconfigurability capability of OBP provides various benefits including flexibility to incorporate future techniques/standards, time-to-market reduction, simplicity of payload structure, flexible business models, and phased array control \cite{5GbookIET}. Although regenerative processing, digital transparent processing and their combination can enable OBP in satellite systems, current OBP functionalities are mainly limited to onboard switching, multiplexing, and traffic routing, and narrowband communications.  In this regard, some important future research directions include the investigation of low-complexity signal processing and machine learning algorithms to enable onboard interference detection and mitigation, localization, spectrum monitoring, BF, precoding, and flexible connections to the inter-satellite links, and analyzing the feasibility of OBP in wideband communications. 
\subsubsection{Integration with 5G and beyond} 5G and beyond networks are expected to support heterogeneous services, and to provide ultra-reliable and low-latency communications links to the massive number of cellular and machine-type/IoT devices. As terrestrial only solutions are not sufficient to support such a heterogeneous and ubiquitous network, the trend is to integrate 5G and beyond networks with the extra-terrestrial networks including satellites. However, there are several challenges in incorporating satellites in 5G and beyond networks, including higher delay with GSO satellites, Doppler shifts in NGSO constellations, scarcity of radio resources and impairments associated with SatCom transceivers and channels. From the standardization perspective, 3GPP activities on the integration of satellites in 5G networks are in very early stage. In this regard, for the seamless integration of satellite and terrestrial networks, future research should investigate novel air interfaces, SDN, NFV and slicing techniques, integrated signaling, OBP, optical feeder links, multicasting, edge caching and intelligent signal-processing techniques to counteract channel and transceiver impairments. Also, how to encourage SatCom and terrestrial operators to integrate their services is an important non-technical future challenge.
\subsubsection{Models for inter-satellite links (ISLs)}
ISLs can enable an NGSO satellite to transmit its contents to adjacent satellites in visibility with the ground stations. Some possible solutions to implement ISLs are to establish a dedicated radio link between NGSO satellites in the same orbit or between NGSO and GSO satellites with the GSO satellite as a relay node. The main challenges for the NGSO-NGSO ISLs are the reduced achievable link budget, especially in the scenarios where satellites in the ISL do not fall in the same orbital plane, and disruptive effects on the radio link (e.g. Doppler, pointing error) caused due to dynamicity of the scenario. For the NGSO-GSO ISL, the main issues include the delay introduced by the relaying via a GSO satellite, and the co-channel interference due to GSO-NGSO frequency sharing, leading to the need of latency minimization and coordinated interference mitigation techniques. One emerging solution for capacity enhancement of ISLs is to employ optical ISLs, however, in the NGSO-NGSO ISL case, the constellation dynamicity makes the pointing even more difficult than in the radio link case. Some promising techniques to be applied for optical ISLs are wavelength-division multiplexing and polarization interleaving. 
\subsubsection{LEO Mega-constellations}
Although most manufacturing problems in the design of LEO Mega-constellations have been successfully overcome and several companies (SpaceX, Amazon, oneWeb, TeleSAT) have already announced their large LEO plans, there are still several challenges to be addressed. First, an accurate coordination is required to avoid collisions that can create a devastating cascade effect, which may lead to the destruction of very large number of satellites.
Secondly, since these Mega-constellations require a large bandwidth, spectrum coordination or spectrum re-utilization techniques will play a fundamental role. Also, despite the use of ISLs, the ground segment must be restructured as well to efficiently use such a Mega-constellation. In particular,  a large number of ground stations located in several locations, will be required to fully exploit the potential of a massive number of satellites. In addition, new protocols for managing the handover of different satellites between GWs must be adopted, and a regulatory challenge in accessing the usable available spectrum or being frequency agile need to be addressed.
\subsubsection{Precoding models and impairments}
The main challenges for the application of precoding in the SatCom scenarios are briefly highlighted hereafter. First, for effective precoding, a perfect synchronization both in time and frequency between different transmitted streams is required. The frequency synchronization is guaranteed when different beams of a satellite use the same clock reference. If this is not the case, some frequency compensation techniques must be considered at the GW side to guarantee perfect frequency synchronization between the beams. To guarantee the timing synchronization, a calibration phase to compensate between the different paths' lengths inside the satellite may be required \cite{ANDRENACCI2016}. Another aspect to be considered when applying precoding is the complexity. Performing the precoding matrix calculation and its application on the Tx streams may require a significant amount of computational resources, thus requiring an update of the GW hardware, especially when precoding has to be employed over a large number of beams. 
\vspace{-10 pt}
\section{Conclusions}
\vspace{-5 pt}
\label{sec: sec5}
SatCom systems can complement terrestrial systems in various emerging use-cases targeted by 5G and beyond networks. Considering the main differences in the design aspects of next generation HTSs from those of terrestrial systems, this letter provided an overview of system modelling and design aspects of next generation HTS systems. Mainly, communication models of HTSs and design aspects of SatCom transceivers were reviewed and discussed while considering the features of SatCom systems. Finally, some open research topics related to OBP, integration of satellite with 5G and beyond networks, ISLs, Mega-LEO constellations and precoding models were discussed.

\vspace{-10 pt}

\footnotesize{
}

\end{document}